\documentclass[12pt]{iopart}

\usepackage{graphicx} 
\usepackage{dcolumn} 
\usepackage{bm} 

\def\sqrtsNN{\mbox{$\sqrt{s_\mathrm{NN}}$}}

\def\GeVc{\mbox{$\mathrm{GeV}/c$}}

\def\lt{\mbox{$<$}}

\newcommand{ \be }{\begin{equation}}    
\newcommand{ \ee }{\end{equation}}    
\newcommand{ \bea }{\begin{eqnarray}}    
\newcommand{ \eea }{\end{eqnarray}}    
\newcommand{ \la }{\langle}    
\newcommand{ \ra }{\rangle}

\usepackage{color}

\begin{document}       
       
  
\title{    
Azimuthal anisotropy: the higher harmonics }


\author{Arthur M. Poskanzer for the STAR Collaboration
\footnote[3]{For the full author list and acknowledgments see 
Appendix "Collaborations" in this volume.}}
\address{MS70R319, Lawrence Berkeley National Laboratory,
Berkeley, CA 94720, AMPoskanzer@LBL.gov}


\date{\today}

\begin{abstract}
We report the first observations of the fourth harmonic ($v_4$) in the
azimuthal distribution of particles at RHIC. The measurement was done
taking advantage of the large elliptic flow generated at RHIC. The
integrated $v_4$ is about a factor of 10 smaller than $v_2$. For the
sixth ($v_6$) and eighth ($v_8$) harmonics upper limits on the
magnitudes are reported.
\end{abstract}



Anisotropic flow, an anisotropy of the particle azimuthal distribution
in momentum space with respect to the reaction plane, is a sensitive
tool in the quest for the quark-gluon plasma and the understanding of
bulk properties of the system created in ultrarelativistic nuclear
collisions. It is commonly studied by measuring the Fourier harmonics
($v_n$) of this distribution~\cite{Methods}.  Elliptic flow, $v_2$, is
well studied at RHIC and is thought to reflect conditions from the
early time of the collision.  Recently, Kolb~\cite{Kolb_v4} reported
that the magnitude and even the sign of $v_4$ are more sensitive than
$v_2$ to initial conditions in the hydrodynamic calculations. Besides
one early measurement at the AGS~\cite{E877PRL94}, reports of higher
harmonics have not previously been published. Some of the present work
has already appeared~\cite{PRL}.

{\it Experiment---} The data come from the reaction Au~+~Au at
$\sqrtsNN = 200$ GeV. The STAR detector main time projection chamber
(TPC) was used in the analysis of two million events. The main TPC
covered pseudorapidity ($\eta$) from --1.2 to 1.2 and the low
transverse momentum ($p_t$) cutoff was 0.15 \GeVc. In the present work
all charged particles were analyzed, regardless of their particle
type. The errors presented in the figures are statistical.

{\it Analysis---} The difficulty is that the signal is small and the
non-flow contribution to the two-particle azimuthal correlations can
be larger than the correlations due to flow.  To suppress the non-flow
effects the current analysis uses the knowledge about the reaction
plane derived from the large elliptic flow.  One method for
eliminating the non-flow contribution in a case when the reaction
plane is known was proposed in~\cite{Methods}. Results obtained with
this method we designate by $v_4\{EP_2\}$. The analysis for $v_4$ was
also done with three-particle cumulants~\cite{cumulants} by measuring
$\la \cos(2\phi_a+2\phi_b-4\phi_c) \ra$.

\begin{figure}[ht]
\resizebox{
\textwidth}{!}{
\includegraphics{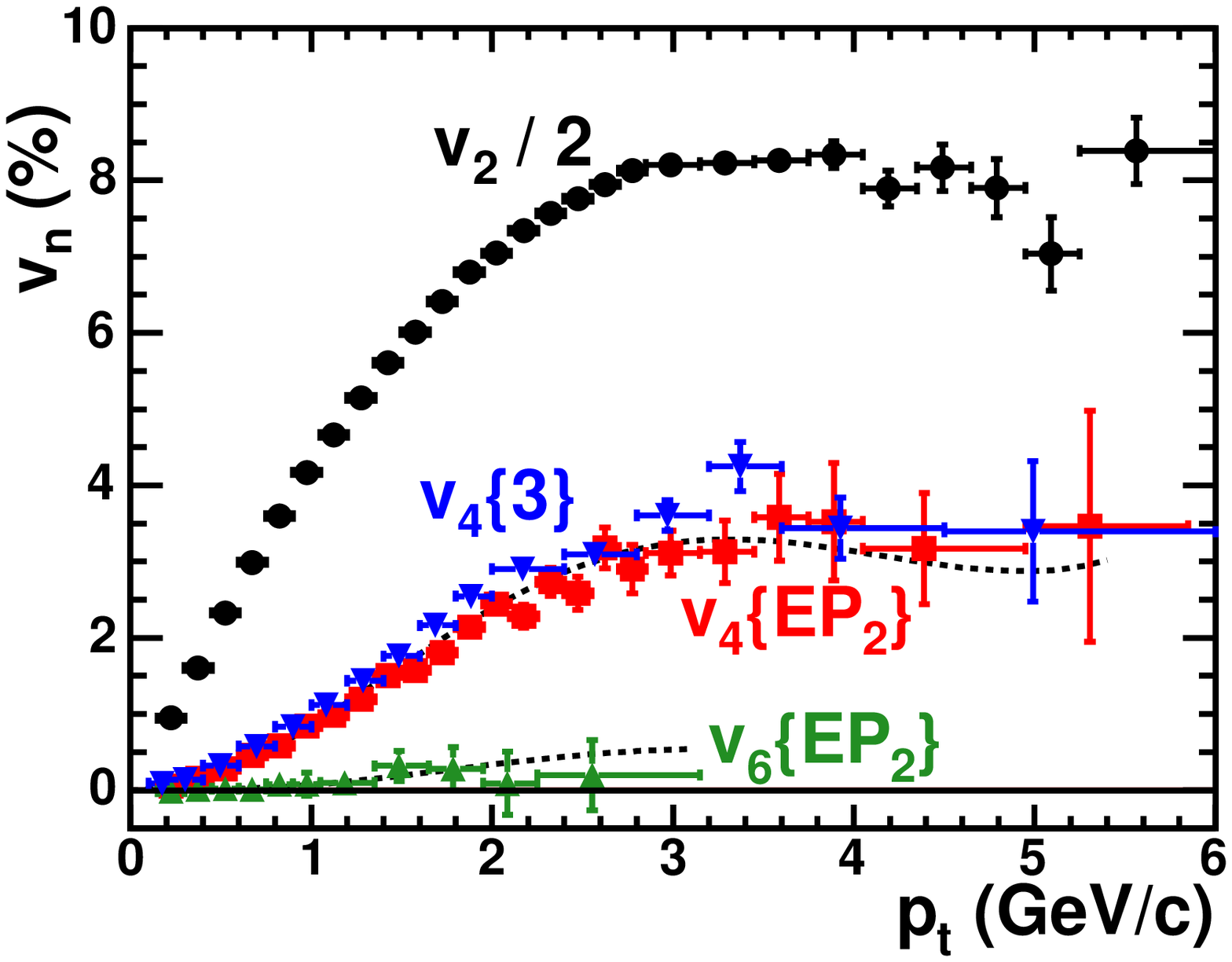}
\includegraphics{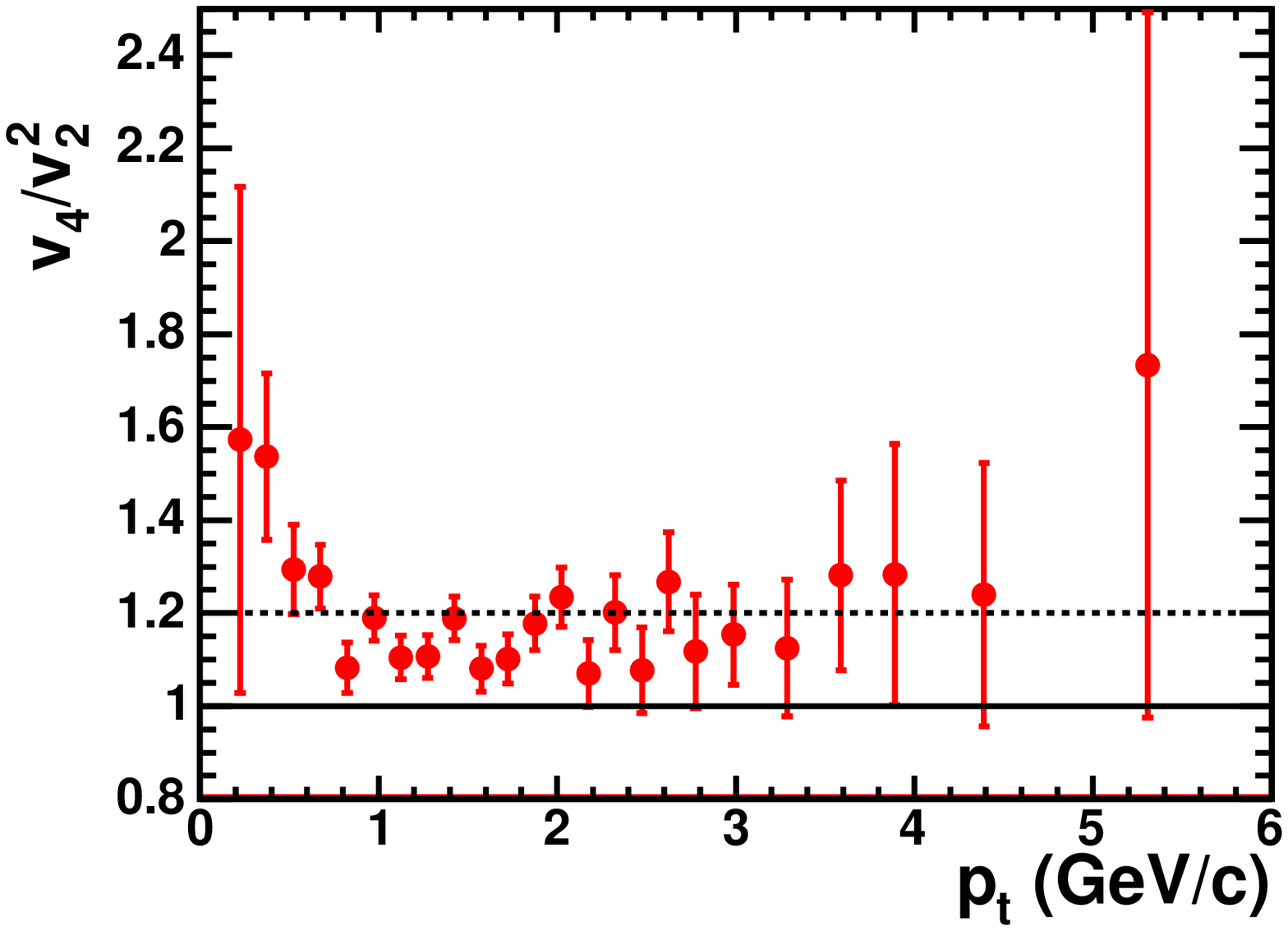}}
\caption{
(left) The minimum bias values of $v_2$, $v_4$, and $v_6$
with respect to the second harmonic event plane as a function of $p_t$
for $| \eta | \lt 1.2$. The $v_2$ values have been divided by a factor
of two to fit on scale. Also shown are the three particle cumulant
values (triangles) for $v_4$ ($v_4\{3\}$). The dashed curves are $1.2
\cdot v_2^2$ and $1.2 \cdot v_2^3$. (right) The ratio $v_4/v_2^2$
is plotted against $p_t$. The dashed line is at the value of 1.2.
\label{v(pt)}}
\end{figure}

{\it $P_t$-dependence---} The results as a function of $p_t$ are shown
in Fig.~\ref{v(pt)} (left) for minimum bias collisions ($0 - 80$\%
centrality). Shown for $v_4$ are both the analysis relative to the
second harmonic event plane, $v_4\{EP_2\}$, and the three-particle
cumulant, $v_4\{3\}$. Both methods determine the sign of $v_4$ to be
positive. As a function of $p_t$, $v_4$ rises more slowly from the
origin than $v_2$, but does flatten out at high $p_t$ like $v_2$. The
$v_6(p_t)$ values are consistent with zero. Ollitrault has
proposed~\cite{OlliScaling} for the higher harmonics that $v_n$ might
be proportional to $v_2^{n/2}$ if the $\phi$ distribution is a smooth,
slowly varying function of $\cos(2 \phi)$. In order to test the
applicability of this $v_2$ scaling we have also plotted $v_2^2$ and
$v_2^3$ in the figure as dashed lines. The proportionality constant
has been taken to be $1.2$ in order to fit the $v_4$ data. The ratio,
$v_4/v_2^2$, is shown in Fig.~\ref{v(pt)} (right) as a function of
$p_t$.

{\it Parton coalescence---} Assuming a simple parton coalescence
model, for mesons one gets~\cite{coalescence}
\begin{equation}
  v_4 / v_2^2 \approx 1/4 + 1/2 (v_4^q/(v_2^q)^2).
\end{equation}
Since experimentally this ratio is 1.2, $v_4^q$ must be greater than
zero. If one assumes that the hadronic $v_2^2$ scaling results from
partonic $v_2^2$ scaling~\cite{partonScaling}, then
\begin{equation}
  v_4^q = (v_2^q)^2
\end{equation}
and 
\begin{equation}
  v_4 / v_2^2 = 1/4 + 1/2 = 3/4.
\end{equation}
But this is still less than 1.2. Therefore either $v_4^q$ is even
greater than simple parton $v_2^2$ scaling would indicate, or the simple
parton coalescence model is inadequate.

{\it Waist---} Kolb~\cite{Kolb_v4} points out that for $v_2 > 10\%$,
which occurs at high $p_t$, and no other harmonics, the azimuthal
distribution is not elliptic, but becomes ``peanut'' shaped.  He
calculates the amount of $v_4$ (which looks like a four-leaf clover)
needed to eliminate this waist.  Our values of $v_4$ as a function of
$p_t$ are about a factor of two larger than needed to just eliminate
the waist.

\begin{figure}[ht]
\resizebox{\textwidth}{!}{
\scalebox{0.7}{\includegraphics{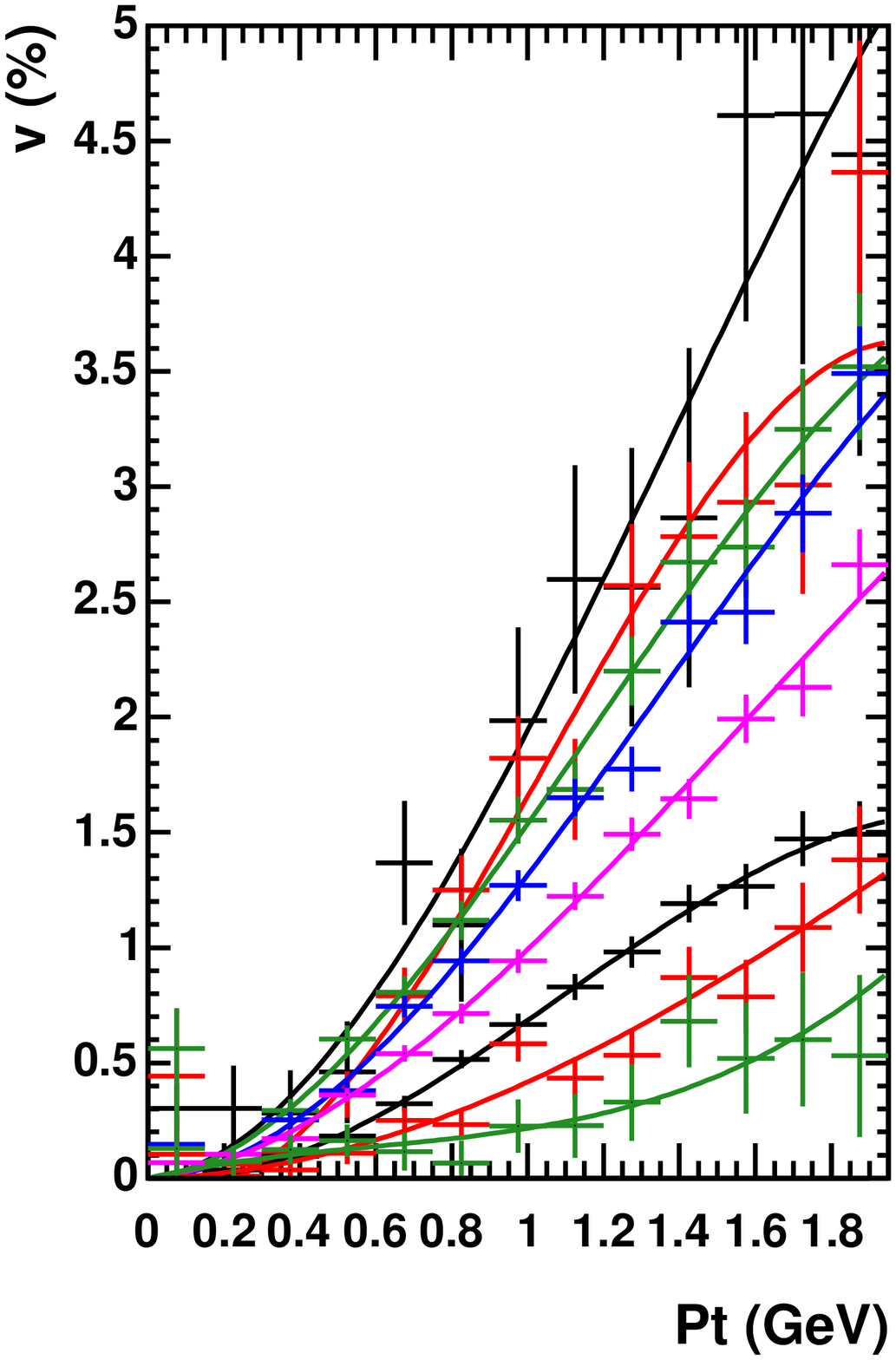}}
\includegraphics{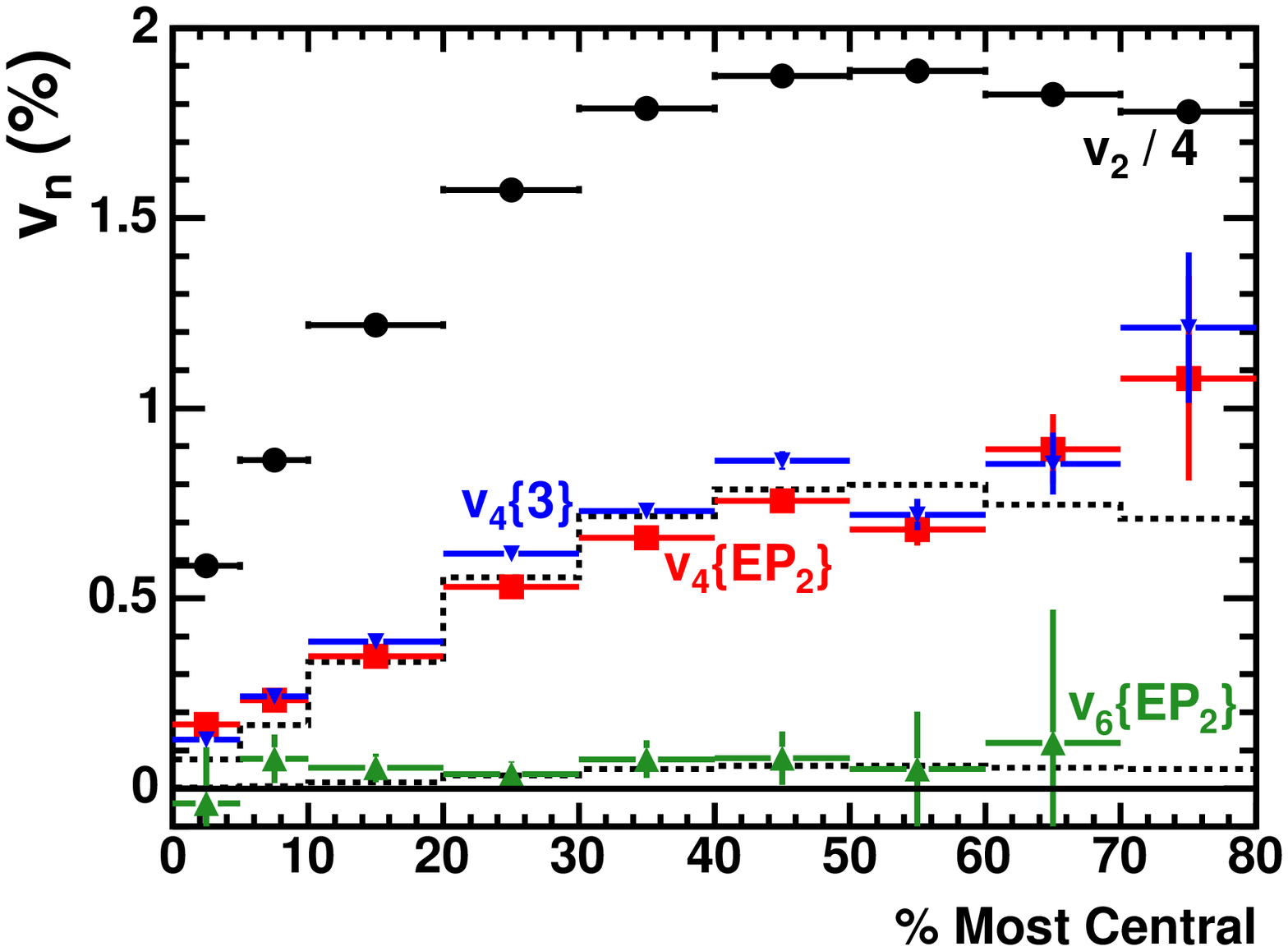}}
\caption{
(left) $v(p_t)$ for the centrality bins (bottom to top) 5 to 10 \% and
10, 20, 30, 40, 50, 60, and 70 up to 80 \%. (right) The $p_t$- and
$\eta$- integrated values of $v_2$, $v_4$, and $v_6$ as a function of
centrality. The $v_2$ values have been divided by a factor of four to
fit on scale. Also shown are the three particle cumulant values for
$v_4$ ($v_4\{3\}$). The dotted histograms are $1.4
\cdot v_2^2$ and $1.4 \cdot v_2^3$.
\label{v}}
\end{figure}

{\it Centrality-dependence---} The values of $v_4(p_t)$ for eight
centrality bins are shown in Fig.~\ref{v} (left). Integrating these
values weighted with the yield gives Fig.~\ref{v} (right) which shows
the centrality dependence of $v_2$, $v_4$, and $v_6$ with respect to
the second harmonic event plane and also $v_4$ from three-particle
cumulants ($v_4\{3\}$). The $v_6$ values are close to zero for all
centralities. To again test the applicability of $v_2^{n/2}$ scaling
we have also plotted $v_2^2$ and $v_2^3$ in the figure as dotted
histograms. The proportionality constant has been taken to be $1.4$ to
approximately fit the $v_4$ data. The larger constant here compared to
that used in Fig.~\ref{v(pt)} is understood as coming from the use of
the square of the average instead of the average of the square, and
because the integrated values weighted by yield emphasize low $p_t$,
where the best factor is slightly larger.

The $v_n\{EP_2\}$ values averaged over $p_t$ and $\eta$ ($ |
\eta | \lt 1.2 $), and also centrality (minimum bias, $0 - 80 \%$),
are (in percent) $v_2 = 5.18 \pm 0.01$, $v_4 = 0.44 \pm 0.01$, $v_6 =
0.043 \pm 0.037$, and $v_8 = - 0.06 \pm 0.14$. Since $v_6$ is
essentially zero, we place a two sigma upper limit on $v_6$ of
0.1\%. Also, $v_8$ is zero, but the error is larger because the
sensitivity decreases as the harmonic order increases.

{\it Blast Wave fits---} We have fitted the data with a modified Blast
Wave model~\cite{PID}:
\begin{equation}
\rho(\phi) = \rho_0 (1 + 2 f_2 \cos(2 \phi) + 2 f_4 \cos(4 \phi))
\end{equation}
\begin{equation} 
  v_n(p_t)= 
  \frac{ \int_{-\pi}^{\pi} d\phi  
    \cos(n \phi) 
    I_n(\alpha_t) K_1 (\beta_t) 
    (1 + 2 s_2 \cos(2\phi) + 2 s_4 \cos(4\phi))} 
  { \int_{-\pi}^{\pi} d\phi 
    I_0(\alpha_t) K_1 (\beta_t) 
    (1 + 2 s_2 \cos(2\phi) + 2 s_4 \cos(4\phi))}, 
  \label{modblastwave}
\end{equation} 
where $I_n$ and $K_1$ are modified Bessel functions, and
$\alpha_t(\phi)=(p_t/T)\sinh(\rho(\phi))$ and
$\beta_t(\phi)=(m_t/T)\cosh(\rho(\phi))$. In these equations, $\rho_0$
is the transverse expansion rapidity ($v_0 = \tanh(\rho_0)$) of the
cylindrical shell. The parameters $f_2$ and $f_4$ are the harmonic
amplitudes of the azimuthal variation of $\rho$, and $s_2$ and $s_4$
describe the spatial anisotropy of the source.

\begin{figure}[ht]
\resizebox{\textwidth}{!}{
\includegraphics{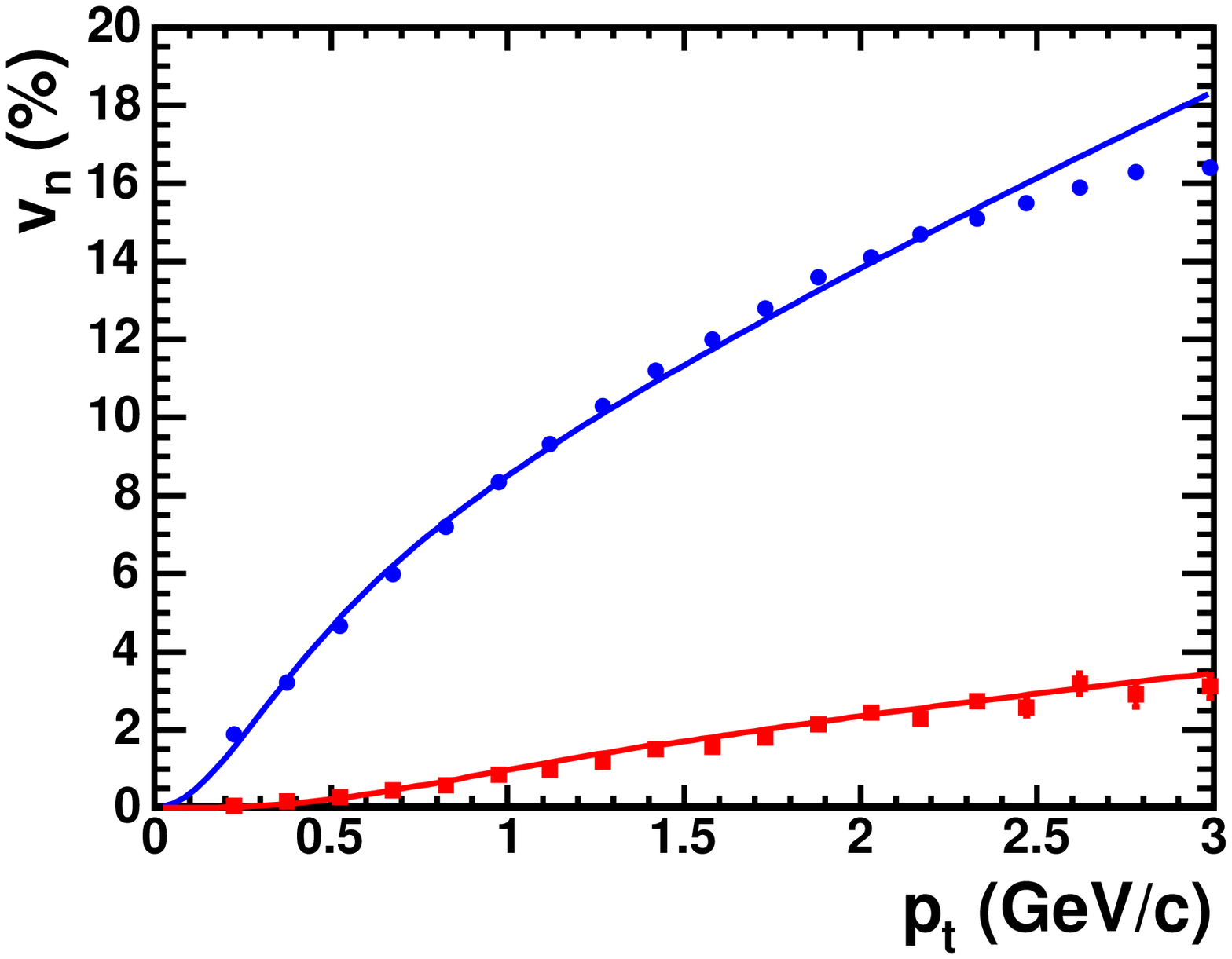}
\includegraphics{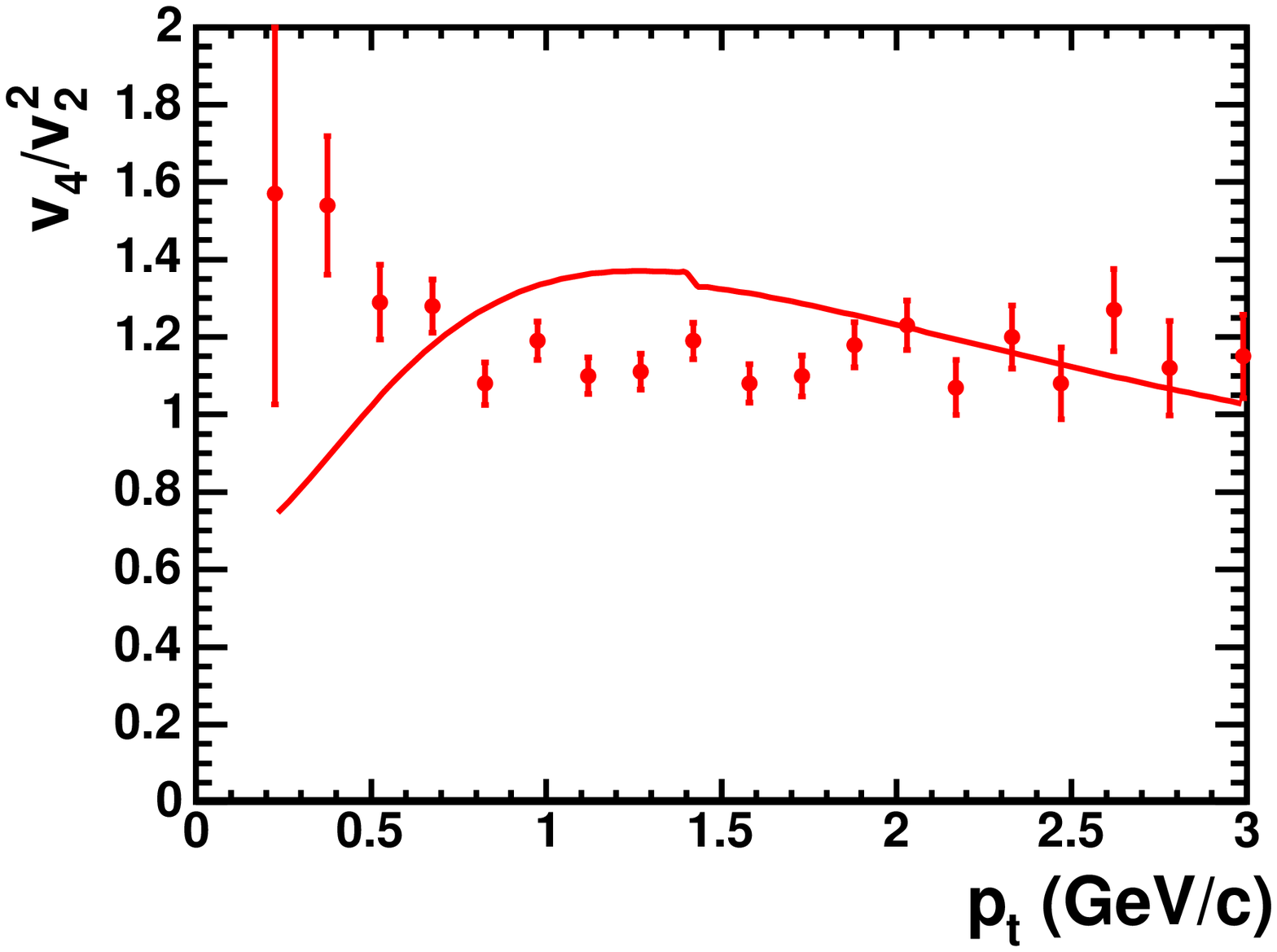}}
\caption{
(left) $v_2$ and $v_4$ as a function of $p_t$ with the lines showing
the Blast Wave fits. (right) The ratio $v_4/v_2^2$ as a function of
$p_t$ with the line showing the ratio of the Blast Wave fits.
\label{BW}}
\end{figure}

The Blast Wave fits to $v_2$ and $v_4$ are shown in Fig.~\ref{BW}
(left) and expanded in Fig.~\ref{BW} (right), showing the approximate
agreement with the ratio. A temperature of 0.1 GeV was assumed giving
the fit parameters $\rho_0 = 0.49$, $f_2 = 1.4 \%$, $s_2 = 9.1 \%$,
$f_4 = 0.0 \%$, and $s_4 = 4.4 \%$. It is interesting that in this
large $p_t$ range the $s$ values are considerably larger than the $f$
values.
 
{\it Conclusions---} We have measured $v_4$ as a function of $p_t$,
and centrality. This is the first measurement of higher harmonics at
RHIC.  It is expected that these higher harmonics will be a sensitive
test of the initial configuration of the system, since they provide a
Fourier analysis of the shape in momentum space which can be related
back to the initial shape in configuration space.


\end{document}